\documentclass{aa}

\usepackage{graphicx}
\usepackage{epsfig}

\newcommand{\nature}[2]{Nature, #1, #2}

\begin{document}
\title{Asymmetry of the PLANCK antenna beam shape and its
manifestation in the CMB data}
\subtitle{}
\author{L.-Y.~Chiang\inst{1} \and P.~R.~Christensen\inst{1,2} \and
H.~E.~J\o rgensen\inst{3} \and I.~P.~Naselsky\inst{4} \and
P.~D.~Naselsky\inst{1,4} \and D.~I.~Novikov\inst{5,6}  \and
I.~D.~Novikov\inst{1,3,6,7}} 
\institute{Theoretical Astrophysics Center, Juliane Maries Vej 30,
DK-2100,  Copenhagen, Denmark
\and  Niels Bohr Institute, Blegdamsvej 17, 
DK-2100 Copenhagen, Denmark
\and University Observatory, Juliane Maries Vej 30, DK-2100,
Copenhagen 
\and Rostov State University, Zorge 5, 344090 Rostov-Don, Russia
\and Astronomy Department, University of Oxford, NAPL, Keble
Road, Oxford OX1 3RH, UK
\and Astro-Space Center of Lebedev Physical Institute,
Profsoyuznaya 84/32, Moscow, Russia
\and NORDITA, Blegdamsvej 17, DK-2100, Copenhagen, Denmark}

\date{Received / Accepted}

\thesaurus{12(12.03.1; 12.04.2) 03(03.19.2; 03.20.9; 03.13.2) 07(07.19.2)}
\maketitle

\markboth{L.-Y. Chiang et~al.:Asymmetry of the {\sc planck} antenna
beam shape and its manifestation in the CMB data }
{L.-Y. Chiang et~al.:Asymmetry of the {\sc planck} antenna
beam shape and its manifestation in the CMB data }
\begin{abstract}
We present a new method to extract the beam shape
incorporated in the pixelized map of CMB experiments. This method is
based on the interplay of the amplitudes and phases of the signal and
instrumental noise. By adding controlled white noise onto the map, the
phases are perturbed in such a way that the beam shape manifests
itself through the mean-squared value of the difference between
original and perturbed phases. This method is useful in extracting
preliminary antenna beam shape without time-consuming spherical
harmonic computations.

\end{abstract}
\section{Introduction}
The future space mission  {\sc planck} will be able to measure, with
unprecedented angular resolution and sensitivity, the cosmic microwave
background (CMB) anisotropy and polarization at 9 frequencies in the
 range 30--857 GHz. In combination with
balloon--borne experiments such as {\sc boomerang},
{\sc maxima}-1,
{\sc tophat} and the recently launched space mission {\sc map}, these
observational data will provide a unique base for investigation of the
history and the large--scale structure formation of the Universe.

The accuracy of the cosmological parameter extraction
planned for the {\sc planck} mission is determined by the
corresponding accuracy of the systematic effects. Systematic
errors can be one of the most important sources of errors for 
high multipole range of the $C(l)$ power spectrum (\cite{mandolesi}).
It is well known that extraction of the information about cosmological
parameters such as baryonic density $\Omega_b$, cold dark matter density
$\Omega_{cdm}$, Hubble constant $H_0$, and so on, needs additional
information about the statistical characteristics of the measured CMB 
anisotropy signal from the sky. The pure CMB signal 
is assumed to be a realization of a random Gaussian
signal on the sphere with power spectrum $C(l)$. The Gaussianity of
the CMB signal means that all its statistical properties are specified
by its power spectrum $C(l)$, which depend on $l$ and not on the
phases. 

In the framework of the CMB observations the signal
 measured by different instruments at different frequencies, however,
displays some peculiarities in observational as well as in foreground
manifestations. This is why a variety of the methods of the
correct information extraction from the CMB data sets are now under 
discussion. All these methods are somewhat complementary to each other in the future highly sensitive CMB experiments, due to different
sensitivity of the methods to different characteristics of the signal.

From a theoretical point of view, the power spectrum of the true CMB
signal is independent of Fourier rings,
meaning that it does not depend on the azimuthal number
$m$. For a flat patch of the sky it corresponds to homogeneity and
isotropy of the signal, without angular dependency of the power spectrum 
$C({\bf k})$ on $\theta=\tan^{-1}({k_y}/{k_x})$,
where ${\bf k}=(k_x,k_y)$. In reality, the CMB signal from the sky has
a more complicated structure reflecting some artifacts of the observation
and different kinds of foreground contaminations, which can
destroy the isotropy of the power spectrum. We will focus on a few
important sources causing artificial anisotropy of the map: (i)
``non-Gaussianity'' (inhomogeneity and anisotropy) of the foregrounds
in the map; (ii) asymmetry of the beam shape, which is now the
standard part of investigation on systematic effects; (iii)
correlations of the instrumental (pixel) noise; (iv) low multipole
modes, e.g. $l \sim 2-10$ for the whole sky ($k\Theta \simeq 1$, where
$\Theta$ is the linear size of the path of the sky), which are
statistically peculiar. Some of the above--mentioned sources of the
$C(l)$ anisotropies are frequency--dependent, thus their contributions
to the maps at different frequency channels of the {\sc planck} are
different. For example, the foregrounds such as dust emission, synchrotron, 
free--free, as well as bright and faint point sources, have different
frequency dependencies and intensities at 33 and 857 GHz range,
which definitely can manifest as some sources of errors in the pixel--pixel
window function and in the corresponding correlation function of the
signals. The influence of the low multipole modes (point (iv)) on the 
possible anisotropies of the maps in the flat sky approximation can be 
detected directly from the corresponding $C({\bf k})$ amplitudes of
the power spectrum.

This paper is mainly devoted to illustration of the idea about 
manifestation and estimation of the asymmetric (elliptical or more
irregular) beam shapes incorporated in the pixelized data, using
both analysis of the two-dimensional spectrum and of phases
 of the map (\cite{naselsky}). We concentrate on beam asymmetry
estimation using simulated CMB map, which reflects directly the
specific of the {\sc planck} scan strategy, map making and noise level. 

There are some important issues related to the beam profiles of the
antenna for the {\sc planck} Low Frequency Instrument (LFI) and High Frequency
Instrument (HFI) frequency channels (\cite{mandolesi}). For instance, down 
to the level -10 dB at LFI, the antenna shapes have approximately
elliptical forms and peculiarities will only be included at higher
multipoles if the level decreases down to -20 dB or less, according to
the design of the Focal Plane Unit (FPU). Thus, roughly
speaking, decentralization of the feed horns in the FPU produces the
optical distortions of the beam shapes from the circular Gaussian
shapes(\cite{burigana}).   

Beam shape influence on the accuracy of the CMB anisotropy $C(l)$ extraction
from the observational data is related to the scanning strategy and
pixelization of the maps from the time-ordered data(TOD)(\cite{wu}). 
During scanning of the CMB sky the antenna beam moves across the sky, meaning
that antenna beam is a function of time. After pixelization of the
TOD the position of each pixel in the CMB map is related
directly with some points in the time stream for which we need to obtain the
information of the orientation of the beam and location of the beam
center relative to each pixel. In principle, given the scanning 
strategy( which for the {\sc planck} mission is under
discussion) and the beam shapes for each frequency channel, we would
be able to model the geometrical properties of the pixel beam shapes and 
their manifestation in the pixel--pixel window functions incorporated 
in the CMB power spectrum $C(l)$. However, the computational 
cost would increase dramatically due to the complicated character of the
pixel--pixel beam matrix(\cite{maino}). Moreover, the scanning strategy 
and the instrumental noise combined with the systematic effects
could transform the actual beam shape during the time of observation.
We then should find some peculiarities of the complicated beam
shape influence on the CMB signal. 

If the response of an antenna on the measured signal is linear and the
CMB signal and the instrumental noise are Gaussian and not correlated,
then the information about the beam anisotropy obtaining by both
methods (power spectrum analysis and phase analysis) is the
same. However in the general case the sets of encoded information obtained
by both methods are different. Thus using both methods is desirable.

The plan of this paper is as follows. In Section 2 we give some
definitions of CMB signals and discuss the basic model of the {\sc
planck} sky map. In Section 3 we introduce a general power spectrum and
phase analysis of CMB signal and the concept of beam--shape
extraction. In Section 4 we describe the main idea and its analytical
approach. The numerical results are presented in Section 5 and the 
conclusion in Section 6. 

\section{Model of the PLANCK sky map}

Let us introduce the standard model of CMB experiment where TOD contain the information about the signal (and noise ) from a
large
numbers of the circular scans. We suppose for simplicity
that all systematic errors are removed after a preliminary ``cleaning''
of the scans. In the temporal domain the observed signal ${m}_t$ is
the 
combination of the CMB + foreground signal ${d}_t$ and random instrumental 
noise $n_t$,
\begin{equation}
m_t=d_t + n_t,
\label{eq:eq1}
\end{equation}
where 
\begin{equation}
d_t = \sum\limits_{l=0}^{\infty} \sum\limits_{m=-l}^{l}B_{t,lm} a_{lm}
Y_{lm}(\vec{x}_t),  
\label{eq:eq2}
\end{equation}
with $B_{t,lm} $ being the multipole expansion of the time--stream beam 
$B_t(\vec{x}_t)$. In Eq.~(\ref{eq:eq2}) $a_{lm}$ is the corresponding
multipole coefficient of the CMB + foreground signal expansion on the 
sphere and $Y_{lm}$ is the spherical harmonics. 
Following Tegmark (1996) we will assume that map--making algorithm is linear. 
The signal in each pixel $s_p$ is then
\begin{equation}
d_t= \sum\limits_{p=0}^{N}{\bf M} _{t,p} s_p,
\label{eq:eq3}
\end{equation}
where $M_{t,p}$ is the corresponding pointing matrix and $s_p$ represents 
the CMB plus foregrounds signals from the sky convolved by the pixel beam 
$B_{p,lm}$,
\begin{equation}
s_p=
\sum\limits_{l=0}^{\infty}\sum\limits_{m=-l}^{l}B_{p,lm}a_{lm}Y_{lm}
(\vec{x}_p) + N_p.
\label{eq:eq4}
\end{equation}
$N_p$ is the instrumental noise in each pixel:
\begin{equation}
N_p=\sum\limits_{l=0}^{\infty}\sum\limits_{m=-l}^{m=l}N_{lm}Y_{lm}
(\vec{x}_p),
\label{eq:eq51}
\end{equation}
and $N_{lm}$ is the coefficient of the pixel noise expansion.
$\vec{x}_p$ is a two--dimensional vector with the components
($x_p,y_p$) to denote the location on the surface of the sphere. The pointing
matrix ${\bf M}_{t,p}$ depends on the scanning strategy of the
observation. Below, as a basic model, we will use the model of the
{\sc planck} mission scanning strategy discussed by Burigana et
al. (2000) with stable orientation of the spin axis in the Ecliptic plane
and without precession of the spin axis. Any precession of the spin
axis will cause additional (regular) spin--axis modulation. 

Instead of the fast rotating beam model described by Wu et al. (2001), we will 
use a stable--orientationed beam model during sky crossing for the spin
and optical axes, i.e., during the rotation of the optical axis around the
spin axis of the satellite, the orientation of the beam is stable with
respect to the optical axis. In addition we will also assume
that instrumental noise is non-correlated for a single
time--ordered scan, and between scans as well. Definitely this model of
instrumental noise is primitive, and needs modifications
and more detailed investigation, but it nevertheless reflects, as shown
in the next section, the geometrical properties of asymmetry of the
beam and their manifestation in the pixelized maps by a given scanning
strategy. Under the assumptions mentioned above we will  use the
elliptical beam shape model of Burigana et al. (2000). For a small
part of the flat sky approximation, we will use the Cartesian
coordinate system with the $x$ axis parallel to the scanning direction
and $y$ axis perpendicular to it. We denote by $x_t$ and $y_t$ the
position of the center of the beam at the moment $t$. Then the beam
shape can be written as  
\begin{equation}
B_t(\vec{x}-\vec{x}_t)= 
\exp\left[-\frac{1}{2}({\bf RU})^T {\bf D}^{-1}({\bf RU})\right],
\label{eq:eq5}
\end{equation}

with

\begin{equation}
{\bf U}=\left(
\begin{array}{c}
        x-x_t \\
        y-y_t

\end{array}
\right),
\end{equation}
where $R$ is the rotation matrix which describes the orientation of
the elliptical beam,
\begin{equation}
{\bf R}=\left(
\begin{array}{rr}
       \cos \alpha  &  \sin \alpha  \\
      -\sin \alpha  &  \cos \alpha 
\end{array}
\right),
\end{equation}
with $\alpha$  the angle between $x$ axis and the principal axis of the
ellipse. The ${\bf D}$-matrix denotes the beam dispersion along the ellipse
principal axis, expressed as
\begin{equation}
{\bf D}=\left(
\begin{array}{cc}
      \sigma^2_{+} &     0          \\
          0        &  \sigma^2_{-}

\end{array}
\right).
\end{equation}
As mentioned by Wu et al. (2001), the signal in the time stream
is not the pixel temperature in $s_p$ itself. The observed signal in
each pixel $p$ depends on the orientation of the pixel beam
$B_p(\vec{x})$ and the location of its center. For a non-symmetric
spatially dependent beam, the convolution of the signal with
asymmetric beam immediately produces asymmetry coupled with the
underlying signal, which affects the estimation of angular power spectrum. 
For beam shape above -20 dB level, where the beam shape is elliptical, we
will use the flat sky approximation in order to demonstrate how we can 
estimate the beam shape in the phases diagram. 
In such an approximation we can describe the signal and instrumental
noise on some small area of the sky using FFT method instead of time consuming
spherical harmonic expansion. In addition we will use the stable pixelized
beam orientation model for a small area of the sky to reflect the scanning
strategy of the observation.

\section{Power and phase analysis of the CMB+noise signal}

Under the assumption of complete sky coverage, the spherical harmonics 
$Y_{lm}(\theta_p,\phi_p)$ can be expressed as a product of the Legendre 
polynomials $P_l^m$ and the signal can be written as (Burigana et al, 1998)

\begin{eqnarray}
s(\theta_p,\phi_p) & = &  \sum \limits_{l=0}^{l_{max}} \sqrt{
  \frac{2l+1}{4\pi} } p_l^0(\cos  \theta_p) {\bf Re}
  (c_{l,0}+N_{l,0}) \nonumber \\ 
 & &  +2 \sum\limits_{m=1}^{l_{max}} \sum\limits_{l=Max\{2,m\}}^{l_{max}}
  \sqrt{\frac{2l+1}{4\pi}} p_l^m(\cos\theta_p) \times  \nonumber\\
& & [ {\bf Re}(c_{lm} + N_{lm}) \cos(m\phi_p)       \nonumber    \\
& &  -{\bf Im}(c_{lm} + N_{lm}) \sin(m\phi_p) ],
\label{eq:eq52}
\end{eqnarray}
where
\begin{equation}
c_{lm}=a_{lm}B_{p,lm},
\end{equation}
and
\begin{equation}
p_l^m(\cos\theta_p)= \sqrt{\frac{(l-m)!}{(l+m)!}}P_l^m(\cos\theta_p).
\end{equation}
We assume for simplicity the {\sc cobe}-like cubic pixelization, which
satisfies the following symmetry properties: if $\theta \in \theta_p$, then
$-\theta \in \theta_p$ also, and if $\phi\in\phi_p$, then 
$\phi+\pi\in\phi_p$.

Using Eq.~(\ref{eq:eq52}) we can introduce the phases $\Psi_{lm}$ of
the signal on the map, by 
\begin{eqnarray}
\tan\Psi_{lm}&=&\frac{{\bf Im}(c_{lm}+N_{lm})}{{\bf Re}(c_{lm}+N_{lm})}=
\nonumber\\
 & = & \frac{|c_{lm}|\sin \Psi_{lm}^{sig} + |N_{lm}| \sin
 \Psi_{lm}^{noise}} {|c_{lm}|\cos \Psi_{lm}^{sig} + |N_{lm}| \cos
 \Psi_{lm}^{noise}},
\label{eq:tangentequation}
\end{eqnarray}
where $\Psi_{lm}^{sig}=\tan^{-1}[{\bf Im}(c_{lm})/{\bf Re}(c_{lm})]$ is 
the phase of the signal from the sky and $\Psi_{lm}^{noise} =
\tan^{-1}[{\bf Im} (N_{lm})/{\bf Re}(N_{lm})]$ is the phase of the
noise. Provided that CMB anisotropy is a random Gaussian field,
the $a_{lm}$ of eq.~\ref{eq:eq2} coefficient is a random variable with
zero mean ($\langle a_{lm} \rangle=0$) and variance $\langle a_{lm} a_{l^{'}m^{'}} \rangle
=\delta_{ll^{'}}\delta_{mm^{'}} C(l)$ where $\delta_{lm}$ is the
standard Kronecker symbol and $C(l)$ the power spectrum. 
Actually, the realization of the random CMB signal on the
sphere is unique, which means that in a $\Delta T(\theta,\phi)$
distribution on the sky we have a single realization of the phases
only. 

We denote the combined signal by $S_{lm}$:
\begin{equation}
S_{lm}=c_{lm}+N_{lm},
\end{equation}
and the power spectrum of this unique realization by $|S_{lm}|^2$. Our
task is to extract the information about the beam shape either using
$\Psi_{lm}$, $|S_{lm}|^2$ or their combination. 

The manifestation of the beam shape asymmetry in the $|S_{lm}|^2$ can
be demonstrated in the following way. Let us consider the following
function

\begin{equation}
\Delta^2_s(lm)=\frac{\langle|N_{lm}|^2\rangle}{|S_{lm}|^2}.
\label{eq:Delta1}
\end{equation}

Qualitatively the properties of the function $\Delta^2_s(lm)$ are the 
following. For small $l$ the CMB signals surpass the noise and the value of $|S_{lm}|^2$ is determined by the 
CMB signal, so for this region $\langle \Delta^2_s(lm) \rangle \ll 1$.
For larger $l$, $|S_{lm}|^2$ becomes smaller. Because 
of the cosmic variance (fluctuations of $|S_{lm}|^2$ as a spectrum of the 
realization of a random process), some $|S_{lm}|^2$ can become close to 
zero, $|S_{lm}|^2<\epsilon^2$ where $\epsilon$ is a constant and  
$\epsilon\ll1$. This means that  $\Delta^2_s(lm)$ at these
points on the $(l,m)$-plane has maxima

\begin{equation}
\Delta^2_s(lm)>\frac{1}{\epsilon^2}.
\label{eq:Delta2}
\end{equation}

\begin{figure}
\centering
\epsfig{file=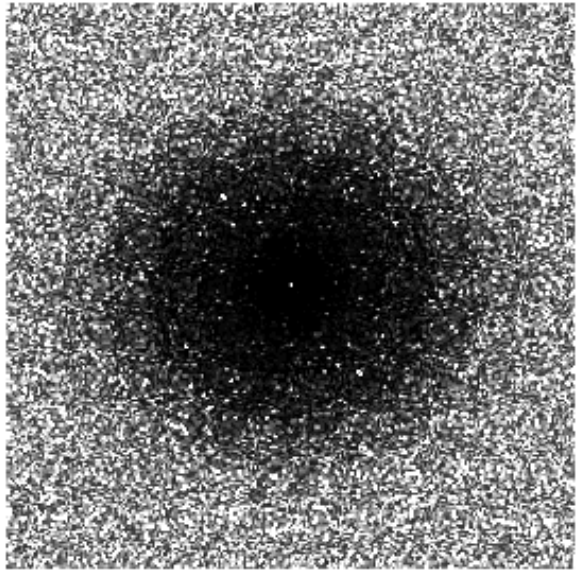,width=8cm}
\caption{Shade-filled contour map of
$\Delta_s(k_x,k_y)=\langle|N_{\bf k}|^2\rangle^{1/2}/|S_{\bf k}|$( in
logarithmic scale). The $x$ and $y$ axis are $k_x$ and $k_y$ axis,
respectively. The detail of the simulation is described in Section
5. For presentation reason( to show the relevant part of the Fourier
ring of $\Delta_s(k_x,k_y)$), we take 25 contour levels between 0 and
$5 \times 10^{-3}$.}\label{deltapower}
\end{figure}

The number density of these maxima increases at larger $l$. For $l$ of 
the order $l_b\sim\Theta^{-1}_b\gg1$, where 
$\Theta_b\sim$~FWHM of the antenna beam, the antenna affects the spectrum.
If the antenna is asymmetric, this influence is also asymmetric. Thus
we have the asymmetric distribution of maxima in this region of the
spectral plane. For very large $l$'s, where the noise $N_{lm}$ dominates
the number density of maxima, Eq.~(\ref{eq:Delta2}) is determined by
the noise, and does not depend on the beam and is therefore
isotropic. The effects are demonstrated in Fig.~\ref{deltapower},
which is a result of a numerical experiment. In order to show the
 beam effect from $\Delta_s(k_x,k_y)$, we add on the symmetric
part. There are 25 contour levels between $\Delta_s(k_x,k_y)=0$ and $5
\times 10^{-3}$. The coordinates represent the flat sky approximation
of the general case at the limit $l,m \gg 1$ (described in Section 5). 

Interval of modes that are sensitive to the beam asymmetry starts from 
$l\sim l_b$ and goes to infinity in ideal conditions (no pixel noise, 
$\langle N^2\rangle=0$). For the real situation the limit of 
this interval is finite and determined by the pixel noise,
$|a_{lm}B_{lm}|^2\sim| N_{lm}|^2$. 

To extract the information about the beam shape from
Fig.~\ref{deltapower} one needs, for example, to draw the averaged
iso-density of the distribution of maxima of $\Delta^2_s$ satisfied to
Eq.~(\ref{eq:Delta2}). One of the possible  methods of drawing this is
described in the next section.

\section{Phase analysis and controlled noise as a probe of the antenna 
          beam shape}

In this section we will show how to estimate the 
antenna beam shape using the information contained in the phase 
distribution of the signal in the map, using a single realization of 
the phases of all $({lm})$ modes.
After the description of this phase method it will be clear that, in the 
simple case of linear response of the antenna, Gaussian CMB signal and 
noise, the phase method is equivalent to the power spectrum method described 
above. However, as we mentioned in the introduction, in the general case these 
methods give different sets of information.

For the phase analysis we will ``perturb'' the phases by adding
controlled white noise into the map. We consider an ensemble $M$ ( $M
\gg 1$), where each element of the ensemble consists of the same
realization of the CMB signal and pixel noise plus a random
realization of a controlled white noise $W_p$ with the variance
$\sigma_W^2=const$ and random phases for each $(lm)$ mode,   
\begin{equation}
s_p^W= s_p + W_p.
\label{eq:addwhitenoise}
\end{equation}
We calculate the ensemble average of the squared difference between phases 
$\Psi^M_{lm}$ of the noise-added realization and the initial one, 
$\Psi_{lm}$:
\begin{equation}
\Delta^2(lm)=\langle \left(\Psi^{M}_{lm}-\Psi_{lm}\right)^2 \rangle|_M
\label{eq:squareddifference}
\end{equation}

The function $\Delta^2(lm)$ from 
Eq.~(\ref{eq:squareddifference}) is considered separately for
different values of the variance of the controlled noise $\sigma_W^2$
in the range from $\sigma_W^2\ll \sigma_N^2$, up to  the  level of
pixel noise $\sigma_W^2\simeq\sigma_N^2$. We will show that the function
$\Delta^2(lm)$ reflects all asymmetric peculiarities of the initial
signal $s_p$. The results of such a kind of phase analysis are tested
numerically and are presented in the next section. Here we describe
the analytical approach to the analysis of the phase mixing effect to
demonstrate how it is possible to reconstruct the antenna beam shape
from the phase distributions in $(lm)$-plane.

\begin{figure}
\epsfig{file=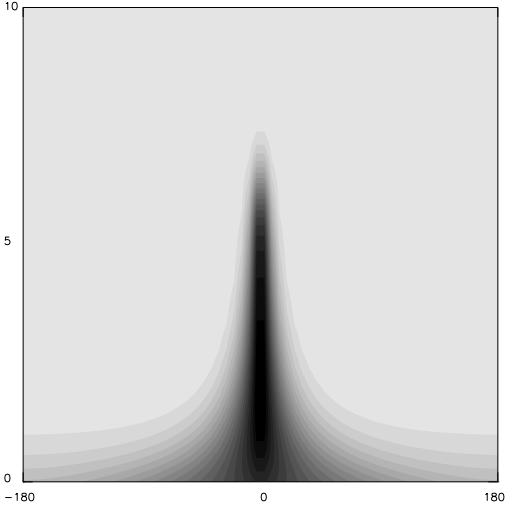,width=7cm}
\caption{The shade-filled map of the integral( Eq.~(\ref{eq:Delta4})) as
a function of two variables: phase $\Psi_{lm}$ along the horizontal
axis and $\rho=|S_{lm}|/\sqrt{\langle|W_{lm}|^2\rangle}$ along the
vertical axis. The intensity of the gray color corresponds to 20 gray
layers from 0 to $\pi^2$.} \label{tabulate}
\end{figure}

For the $s^W_p$ signal the definition of the combined phases $\Psi^{M}_{lm}$ 
for each $(lm)$ mode is similar to Eq.~(\ref{eq:tangentequation}): 
\begin{equation}
\tan \Psi^{M}_{lm}=\frac{|S_{lm}| \sin \Psi_{lm}+|W_{lm}| \sin \Phi^W_{lm}}
{|S_{lm}| \cos\Psi_{lm}+|W_{lm}| \cos\Phi^W_{lm}}
\label{eq:tangent2equation}
\end{equation}
where $S_{lm}$ is the multipole expansion of the combined signal(CMB $\otimes$ 
BEAM +  PIXEL NOISE) at the mode $(lm)$, $\Psi_{lm}$ the corresponding 
phase, $W_{lm}$ the controlled noise expansion, and $\Phi^W_{lm}$ its 
phase. The analytical expression for the function $\Delta^2(lm)$ can be 
written in the following way
\begin{eqnarray}
\lefteqn{\Delta^2(lm)  = \frac{1}{2\pi}
\int_{-\Psi_{lm}}^{2\pi-\Psi_{lm}} \varphi^2 d\varphi} \nonumber \\
&& \left\{ e^{-\rho^2/2}  + \frac{ \sqrt{\pi}} {2} \rho \cos \varphi
\left[ 1 + \Phi \left(\frac{ \rho \cos\varphi}{\sqrt{2} } \right)
\right] \right\},  \label{eq:Delta4}
\end{eqnarray}
where 
\begin{equation}
\rho=\frac{|S_{lm}|}{\sqrt{\langle|W_{lm}|^2\rangle}},
\end{equation}
and
\begin{equation}
\Phi(x)=\frac{2}{\sqrt{\pi}}\int^x_0e^{-t^2}dt.
\end{equation}
The integral in Eq.~(\ref{eq:Delta4}) has been tabulated as a function of 
two variables $\Psi_{lm}$ and $\rho$ and the result is presented in
Fig.~\ref{tabulate}. We will use this tabulation for the numerical
experiment in the next section. In this section we will  obtain simple
analytical asymptotics of Eq.~(\ref{eq:squareddifference}).

From  Eq.~(\ref{eq:Delta4}) one can find the difference
\begin{equation}
\tan \Psi^{M}_{lm}- \tan \Psi_{lm}= \frac{\frac{|W_{lm}|}
{|S_{lm}|} \sin(\Phi^W_{lm}-\Psi_{lm})} 
{\left(1+\frac{|W_{lm}|\cos\Phi^W_{lm}}
{|S_{lm}| \cos\Psi_{lm}}\right) \cos\Psi_{lm}}.
\label{eq:tangentdifference}
\end{equation}
Let us determine the mean--squared value $\Delta^2_{\tan}(lm)$ of the
difference between $\tan \Psi^{M}_{lm}$ and $\tan\Psi_{lm}$ using
 Eq.~(\ref{eq:tangentdifference})
\begin{equation}
\Delta^2_{\tan}(lm)=\frac{1}{2\pi}\int_0^{2\pi}d\Phi^W_{lm}\left
(\tan\Psi^{M}_{lm}-\tan\Psi_{lm}\right)^2.
\label{eq:meansquaredtangent}
\end{equation}
One finds that $\Delta^2_{\tan}(lm) \simeq \Delta^2(lm) /
\cos^4\Psi_{lm}$ ~if~$\Delta^2(lm) \ll 1$ and after integration we obtain
\begin{eqnarray}
\Delta^2_{\tan}(lm) & = & \frac{2\alpha_{lm}^2(1+\alpha_{lm}^2)}
{(1-\alpha_{lm}^2)^3 \cos^2
\Psi_{lm}} \nonumber \\
 & & \times \left[1-\frac{\alpha_{lm}^2(3-\alpha_{lm}^2)}
{1+\alpha_{lm}^2} \cos 2 \Psi_{lm}\right]
\label{eq:meansqtangentsol}
\end{eqnarray}
where $\alpha_{lm}=(1-\sqrt{1-\beta_{lm}^2})/\beta_{lm}$, and 
\begin{equation}
\beta_{lm}=\frac{\sqrt{\langle|W_{lm}|^2\rangle}}{|S_{lm}|} \cos\Psi_{lm}\le \frac{1}{2}.
\end{equation}
In the asymptotic $|\beta_{lm}| \ll 1$ we have
$\alpha_{lm}=\frac{1}{2}\beta_{lm}\ll 1$ and from
Eq.~(\ref{eq:meansquaredtangent}) and Eq.~(\ref{eq:meansqtangentsol})
we get
\begin{equation}
\Delta^2(lm)\simeq \frac{{\langle|W_{lm}|^2\rangle}}{2|S_{lm}|^2}\left[1-\frac{3{\langle|W_{lm}|^2\rangle}}
{4|S_{lm}|^2\cos^2\Psi_{lm}}\cos2\Psi_{lm}\right].
\label{eq:approxmeansquared}
\end{equation}
For our approximation we can neglect the second term in the brackets of 
Eq.~(\ref{eq:approxmeansquared}) and
\begin{equation}
\Delta^2(lm)\approx \frac{{\langle|W_{lm}|^2\rangle}}{2|S_{lm}|^2}.
\label{eq:Delta3}
\end{equation}
Properties of ${\langle|W_{lm}|^2\rangle}$ are the same for any values of $l$ (and $m$). 
Properties of $1/2|S_{lm}|^2$ were discussed in the previous 
section when we discussed $\Delta^2_s(lm)$. It is obvious that qualitatively 
the behaviors of the function $\Delta^2(lm)$ are the same as 
$\Delta^2_s(lm)$.
But there is one principal difference between the qualitative and the full 
correct description of the asymptotic $|S_{lm}|^2\rightarrow 0$. Namely, 
if $|S_{lm}|^2\rightarrow 0$ then
${\langle|W_{lm}|^2\rangle}/|S_{lm}|^2 \gg 1$, and the asymptotic 
Eq.~(\ref{eq:approxmeansquared}) is not valid. 
Under the condition $|W_{lm}|^2\gg |S_{lm}|^2$ we have
$\Psi^M_{lm}\simeq \Phi^W_{lm} $ and from
Eq.~(\ref{eq:squareddifference}) and Eq.~(\ref{eq:tangent2equation})
we reach 
\begin{equation}
\Delta^2(lm)\simeq \frac{4\pi^2}{3}-\pi \Psi_{lm} +\Psi^2_{lm}.
\label{eq:asymptotic2}
\end{equation} 
In fact the controlled white noise is some kind of averaging factor in our
 phase analysis. An analogous method may be used for the power spectrum 
analysis of the beam asymmetry.

\section{The flat sky approximation and numerical results}

The general properties of $|S_{lm}|^{2}$ allow us to introduce more
convenient and  faster flat sky analysis, which is specifically useful 
for the antenna beam shape estimation. This model
reflects the scanning strategy at present specified for the {\sc planck}
mission, when, for a small part of the sky far from the North and
South poles, the model of the stable and fixed beam
orientation is adequate, without rotation and multi-crossing scans. The implementation of FFT significantly decreases the
computational cost, which, for time consuming spherical harmonic
analysis, is a major issue. 

Using definitions of the signals and noises from the previous section, we 
define $S_{sky}({\bf k})$ to be the Fourier component of the signal from the 
sky measured by antenna and $N({\bf k})$ that of the 
noise, which does not depend on the beam properties. The Fourier modulus
and phases of the combined signal are defined as follows,

\begin{equation}
S({\bf k})=S_{sky}({\bf k})+N({\bf k})=\sqrt{C({\bf k})}\exp[i\Phi({\bf k})].
\label{eq:eqf1}
\end{equation}
The function $C({\bf k})$ is the power spectrum of the combined signal 
(observational data and noise) and  $\Phi({\bf k})$ describes the phases. 
$S_{sky}({\bf k})$ represents the CMB signal convolved with the beam. If the 
signal from the sky does not contain non-Gaussian foreground  components (or 
its manifestation is suppressed at the ``cosmologically clean'' channels 
from 70 to 217 GHz (\cite{mandolesi})) and if the noise $N({\bf k})$ is 
uncorrelated white noise, the distribution of the phases $\Phi({\bf k})$ for 
different ${\bf k}$ must be random and uniform. If there are some 
contaminations from the foregrounds and(or) pixel noise is of non-Gaussian 
nature, the phase distribution over the ${\bf k}$ range can be weakly correlated, 
and can be tested by phase diagram method(\cite{cc2}). Such kind of 
uncertainties of the statistical properties of the signal $C({\bf k})$ 
are important for separation of the pure CMB signal and noise for complicated
characters of the beam shape. 

We start out with a squared Gaussian random map with the power spectrum
from the angular power spectrum of the  $\Lambda$CDM model from Lee et
al. (2001). The map simulates a $25.6^\circ \times 25.6^\circ$ square 
realization of the CMB temperature fluctuations with pixel size 3 arcmin and 
periodic boundary conditions (PBC) (see \cite{BE} and remarks in Section 6). 
We then add pixel noise, $\sigma_{noise} \sim 6 \times 10^{-6}$, after 
convolving the map with an elliptical beam with long-axis and short-axis FWHM
12 and 9 arcmin, respectively. After generation of the map, which models
the HFI 100~GHz frequency channel, we sum $M=10^2$ realizations of the
controlled noise with variance $W^2$ close to the pixel noise variance
(Eq.~(\ref{eq:addwhitenoise}) and $W^2 \sim \langle N^2({\bf
k})\rangle$) and calculate the mean squared difference between phase of
the signal and that of the signal plus controlled noise in the flat
sky approximation:

\begin{equation}
\Delta^2({\bf k})=\langle\left[\Psi_n({\bf k})-\Psi({\bf
k})\right]^2\rangle |_M.
\label{eq:eq99}
\end{equation}

\begin{figure}
\centering
\epsfig{file=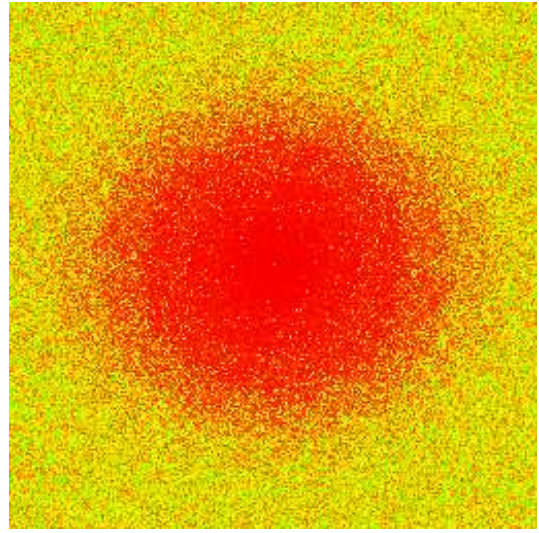,width=7cm} 
\caption{The colored phase map for the $\Delta({\bf k})$ function. The
$x$ and $y$ axis are $k_x$ and $k_y$ axis, respectively. For
presentation reason, we show the symmetric part( Fourier ring) in the
Fourier domain. The size of the phase map is ranged from $-{\bf k}$ to
${\bf k}$ ( $|{\bf k}|=128$), where $2 \pi |{\bf k}|^{-1} \sim $ 12 arcmin. } 
\label{colorphase}
\end{figure}

Analogously we can calculate $\Delta_s(k)$ for the flat
approximation. In Fig.~\ref{deltapower} we display the function
$\Delta_s(k)$, it is rather difficult, however, to draw the
averaged contour lines for $\Delta_s(k)$ directly from this map. It is
easier to do this for the map of the  $\Delta(k)$ function.
Fig.~\ref{colorphase} displays the $\Delta({\bf k})$ function using color
representation of phases (\cite{cc2}). In the color representation, phases
are mapped onto the color circle, so that the phase of each mode is
represented by a color hue. Color red represents $0^\circ$ and the
primary hues are $120^\circ$ apart, i.e., phases of $120^\circ$
and $240^\circ$ are represented by color green and blue. Cyan and
other complementary colors magenta and yellow represents $180^\circ$
and $300^\circ$, and $60^\circ$, respectively. For presentation
effect, we add the symmetric part (Fourier ring) in the Fourier
domain. It is clear from the color map of Fig.~\ref{colorphase} that in the center for large-scale modes( small ${\bf
k}$), the values of $\Delta({\bf k})$ are approaching zero, shown
by color red. This is due to the fact that the phases of large-scale
modes( the inner rings) are not ``perturbed'' by added noise. On small-scale
modes( the outer rings), where the phases are dominated by pixel noise,
the added controlled noise randomizes the resulting phases when the controlled
noise level is chosen the same as that of pixel noise.

\begin{figure}
\epsfig{file=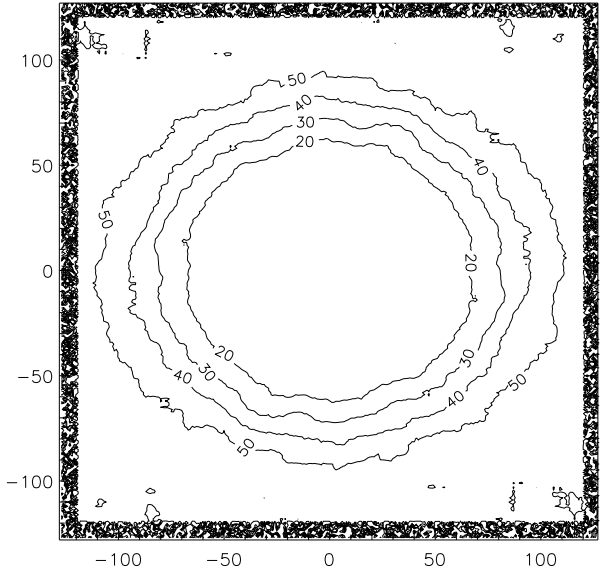,width=9cm}
\caption{The contour map of $\Delta_{sm}({\bf k})$ for $k_s=7$. The
contour level is in unit of degree. The size of the phase map
is taken from $-{\bf k}$ to ${\bf k}$( $|{\bf k}|=128$), where $2 \pi
|{\bf k}|^{-1} \sim$ 12 arcmin.}   
\label{contour}
\end{figure}

In the intermediate regime, the beam shape manifests itself through
the $\Delta({\bf k})$ function. The fuzzy regions in the intermediate regime 
reflect the anisotropy of the beam shape. By taking average as
\begin{equation}
\Delta_{sm}({\bf k})=\frac{1}{(2k_s+1)^2}\sum_{{\bf k}^{'} \in S}
 \Delta({\bf k}^{'}),
\label{eq:smoothing}
\end{equation}  
where $S=\{{\bf k^{'}}:\{|k_x-k_x^{'}|<k_s$, $|k_y-k_y^{'}|<k_s\}\}$,
we can find the mean shape of the beam. Figure \ref{contour} shows the
contour map of the $\Delta_{sm}({\bf k})$ function by $k_s=7$. From the
contour map, the elliptic shape is estimated roughly to have the same ratio
of the simulated beam shape, i.e., 4/3.

\begin{figure}
\centering
\epsfig{file=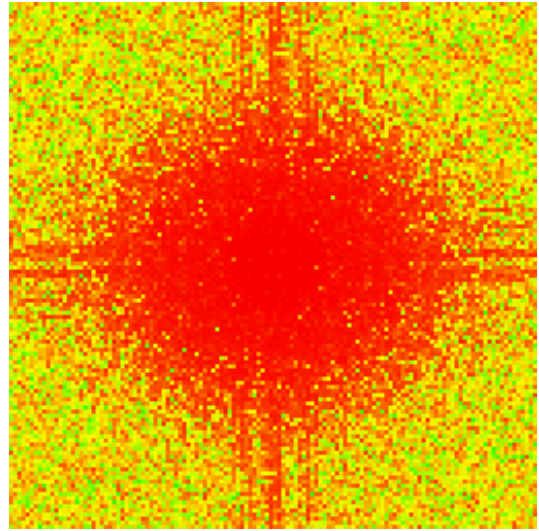,width=7cm} 
\caption{The colored phase map of $\Delta$ function for non-PBC map,
which is extracted from the central part of the previous $25.6^\circ
\times 25.6^\circ$ PBC simulated map. The $x$ and $y$ axis are $k_x$ and
$k_y$ axis, respectively. To compare with Fig.~\ref{colorphase}, we
only show half of the $k$-range of Fig.~\ref{colorphase}, i.e., from
$-{\bf k}$ to ${\bf k}$( $|{\bf k}|=64$), where $2 \pi |{\bf k}|^{-1}
\sim$ 12 arcmin.} 
\label{colorphasenp}
\end{figure}

An important issue is related to the asymmetric beam extraction
from a real map, covering a small patch of the sky. Previously we used
periodic boundary condition (PBC) for modeling the CMB signal
(\cite{BE}). In reality, for some square $\Theta\times\Theta$ patch of
the whole sky map, the PBC is artificial and one can ask how
sensitive this controlled noise method is to the deviation of the
artificial phases of the signal from the true distribution? To answer
this question we show in Fig.~\ref{colorphasenp} the result of a
numerical experiment for a map which was constructed in the
following way: We generate, as described earlier, the PBC map with the size
$\Theta_p\times\Theta_p$, $\Theta_p=25.6^\circ$, and extracted from
this map the inner part $\Theta_{np}\times\Theta_{np}$  with the size
$\Theta_{np}=12.8^\circ$. We then apply the controlled noise
method for the  non-PBC map and compare our results of the beam
extraction with the PBC case. Figure~\ref{colorphasenp} shows the
colored map of this $\Delta$ function. In order to compare with the PBC case,
 only half of $k$-range of interest is extracted from non-PBC case, as
the beam size relative to the map is now twice of the PBC case. The
peculiarity of the red crossing is induced by the non-PBC.  
The difference between these two models is less then $5{\%}$. This
implies that we can apply our method for small real patch of the
sky directly. 

\section{Conclusion}
In this paper we propose the method of power and phase analysis to
demonstrate the manifestation of the  anisotropy of the antenna beam
shape incorporated in the pixelized CMB  anisotropy data, and how to
estimate the shape of the beam.

It is worth noting that the numerical realization of the phase diagram method 
described above is based on the flat sky approximation and illustrates 
the general properties of the CMB signal extraction from the pixelized sky 
map at the high multipole limit $l\gg1$. In such a case, the Fourier analysis
reflects directly the general properties of the phases of a signal on the 
sphere, which in the general (whole sky) limit must be defined using 
spherical harmonic analysis. But there are a few  balloon experiments ({\sc 
boomerang}, {\sc maxima}, {\sc tophat}) which cover relatively small areas of
the sky in comparison with {\sc planck} and the recently launched {\sc map} 
missions. Moreover, for the beam shape extraction from {\sc map} and 
{\sc planck} missions it will be convenient to extract preliminary
information about antenna beam shape without time consuming spherical 
harmonic computations. In connection with the {\sc planck} mission
this approach looks promising due to otherwise high computation cost
in the framework of the $C(l)$ extraction program. 

As it is shown, the functions $\Delta^2_s({\bf k})$ and $\Delta^2({\bf k})$
reflect the general properties of the power spectrum ${C({\bf k})}$
measured from the small patch of the sky.

At the end of this discussion we would like to give the following remark. 
For estimation of the asymmetry of the antenna beam shape using 
the controlled noise method, we need to know the limit of the beam 
cross-level (in dB), for which we can extract the peculiarities of the beam.
According to general prediction of the beam shape properties for the 
{\sc planck} mission it is realistic to assume that the ellipticity of
the beam preserves up to the cross-level $\nu\simeq -10$ dB. For a
small patch of the map we can use $B({\bf k})\sim
\exp\left[-k^2_x/(2\sigma^2_{+})-k^2_y/(2\sigma^2_{-})\right]$ and
$C({\bf k})$ for the best fit $\Lambda$CDM
cosmological model from the MAXIMA-1 data. Because of logarithmic dependence of the beam shape parameters 
on the $C({\bf k})$ and $\sigma^2_{N}$:
$$
\frac{k^2_x}{2\sigma^2_{+}} + \frac{k^2_y}{2\sigma^2_{-}}\simeq -
\ln\left[\frac{C({\bf k})}{\sigma^2_{N}}\right]
$$
we can find the ratio $\sigma^2_{-}/ \sigma^2_{+}$ with 3-5 $\%$ accuracy at
the cross-level $\nu\simeq -6$ dB. This method is most applicable to
the patch of sky at low declination where the assumption of stable
orientation of the antenna beam is satisfied. A more general case will be
investigated in a separate paper.

\section*{Acknowledgment}

This paper was supported in part by Danmarks Grundforskningsfond 
through its support for the establishment of the Theoretical 
Astrophysics Center, by grants RFBR 17625 and INTAS 97-1192.

\end{document}